\documentstyle[preprint,aps]{revtex}

\begin{document}

\preprint{UT/W94-007\\
hep-th/9402026
}

\draft

\title{
On the Additional Symmetry;\\
Many-Body Problem Related to the KP Hierarchy
}

\author{Kazuhiro Hikami and Miki Wadati}

\address{Department of Physics, Faculty  of  Science,\\
  University of Tokyo,\\
  Hongo 7-3-1, Bunkyo-ku, Tokyo 113, Japan}

\date{January 1994}

\maketitle

\begin{abstract}
Nonlinear integrable equations, such as the KdV equation, the
Boussinesq equation and the KP equation,  have the close relation with
many-body problem.
The solutions of such equations are the same as  the restricted flows
of the classical  Calogero model, which is one-dimensional particle
system with inverse square interactions.
The  KP hierarchy and the Calogero model share the same
structure  called  ``additional symmetry''.
This additional symmetry plays  a crucial role  in this relation.

\end{abstract}

\pacs{}

\narrowtext

In 1980s Sato school has developed the soliton theory and established
the relationship with the infinite dimensional Lie
algebra~\cite{DKJM83}.
The basic equation  is the Kadomtsev-Petviashvili (KP)  hierarchy.
Nonlinear integrable equations, such as the Korteweg-de Vries (KdV)
equation and the Boussinesq equation, can be considered as the
reduction of the KP hierarchy.

Well known is  that such nonlinear evolution equations are closely
related not only to the string theory~\cite{BK90}, but to the
one-dimensional many-body problems~\cite{AMM77,OP81}.
The rational solutions of the equations correspond to the (restricted)
flows of the so-called  Calogero model~\cite{CAL69}.
For example,  let us consider the KdV equation,
\begin{equation}
  u_t + 6 u u_x + u_{xxx} = 0 .
    \label{kdv}
\end{equation}
The form of pole-expansion,
\begin{equation}
  u (x,t) = - 2 \sum_{j=1}^N \frac{1}{ ( x- q_j(t) )^2 } ,
\end{equation}
is the solution if $ q_j $ follow the Hamiltonian flow induced by
$ {\cal I}_3 $ (see below) with constraint,
\begin{equation}
  \sum_{j \neq k}^N \frac{1}{ (q_j - q_k)^3 } = 0 , \; \; k = 1, 2,
  \cdots, N .
\end{equation}
The meaning of these relations still remains as an open
problem~\cite{AC91}.

In this letter we give the meaning based on the ``additional
symmetry''.
The additional symmetry of the KP hierarchy has been introduced
in~\cite{OS86,Dic91}.
This symmetry plays an important role  in the matrix model for
two-dimensional gravity,  and  provides a string theory,  Virasoro
constraint, and so on.
On the other hand,  the  symmetry of the Calogero model has also been
revealed for classical \cite{BR77} and quantum \cite{HW93d}
cases.
We show that   the Calogero model has the same structure with the KP
hierarchy.

We begin with the Sato theory and   consider the symmetry of the KP
hierarchy.
Let $ \{ t_j \} $ denote a set of independent variables.
We set  $ x \equiv t_1 $ and $ \partial = \partial/\partial x $.
The formulation   is based on the Lax equations for
the pseudo-differential operator,
$
L = \partial + u_2 \partial^{-2} + u_3 \partial^{-3} + \cdots
$~:
\begin{equation}
  \frac{\partial L}{\partial t_n} = \left[ B_n , L \right] ,
    \; \; n = 1, 2, \cdots .
    \label{lax}
\end{equation}
Here $ u_n $ are the functions of $ \{ t_j \} $, and
$ B_n = ( L^n )_{+} $
is the truncation to  the differential part of $ L^n $.
A negative power part of the differential operator $ L^n $ will be
written as $ ( L^n )_- $.
The Lax equations (\ref{lax}) satisfy  the zero-curvature equation,
which is known as the Zakharov-Shabat equations,
\begin{equation}
  \frac{\partial B_n}{\partial t_m} - \frac{\partial B_m}{\partial t_n}
    + \left[ B_n , B_m \right] = 0 ,
   \; \; n, m = 1, 2, \cdots .
\end{equation}
Note that the Lax equations (\ref{lax}) give the integrability
conditions of the linear system,
\begin{mathletters}
\begin{eqnarray}
  & & L \psi(t,\lambda) = \lambda \psi(t, \lambda) , \\
  & & \frac{\partial \psi(t,\lambda)}{\partial t_n}
  = B_n \psi(t, \lambda) .
\end{eqnarray}
\end{mathletters}
where $ \psi(t,\lambda) $ is the Baker-Akhiezer function.
This function  can be written as
\begin{equation}
  \psi(t,\lambda) = W \exp ( \sum_{n=1}^\infty t_n \lambda^n ) ,
\end{equation}
where $ W $ is the zero-th order pseudo-differential operator
\begin{equation}
  W = 1 + w_1(t) \partial^{-1} + w_2(t) \partial^{-2} + \cdots .
\end{equation}
This dressing operator $ W $ satisfies the Sato equation
\begin{mathletters}
\label{sato}
\begin{eqnarray}
  & &  \frac{\partial W}{\partial t_n} = B_n W - W \partial^n , \\
  & &  L W = W \partial .
\end{eqnarray}
\end{mathletters}
It is  known that the KP hierarchy has the higher symmetry~\cite{OS86}.
This symmetry plays an important role for the theory of the
two-dimensional gravity~\cite{FKN91}.
We define the operator $ M $ by
\begin{equation}
  M = W \Bigl( \sum_{k=1}^\infty k t_k \partial^{k-1} \Bigr) W^{-1}
    = W x W^{-1} + \sum_{k=2}^\infty k t_k L^{k-1} .
\end{equation}
This operator satisfies the Lax equation
\begin{equation}
  \frac{\partial M}{\partial t_n} = \left[ B_n , M \right] .
    \label{add}
\end{equation}
It can be checked that the operators $ L $ and $ M $ are canonically
conjugate,
\begin{equation}
  \left[ L , M \right] = 1 .
    \label{lm}
\end{equation}
By use of these operators we can construct the $ W_{1+\infty} $
algebra \cite{FKN91},
\begin{equation}
  \left[ {\cal W}_n^{(s)} , {\cal W}_m^{(s')} \right]
  = ( (s'-1) n - (s-1) m )
  {\cal W}_{n+m}^{(s+s'-2)} + \cdots ,
  \label{winf}
\end{equation}
with the $ W $-operators,
$
{\cal W}_n^{(s)} =  M^{n} ( M L )^{s-1} .
$
The relation (\ref{lm}) and its generalized form,
\begin{equation}
  [ L^n , \frac{1}{n} M L^{-n+1} ] = 1 ,
\end{equation}
correspond to the string equation in the theory of the
two-dimensional gravity~\cite{BK90}.
Notice that (\ref{add}) and (\ref{lm}) are consistent with  the linear
equation
\begin{equation}
  \frac{\partial \psi(t,\lambda)}{\partial \lambda} = M \psi(t,\lambda) .
\end{equation}
For our purpose it is more convenient to rewrite (\ref{lax}),
(\ref{add})  and (\ref{sato}) into
\begin{mathletters}
\label{lax2}
\begin{eqnarray}
  & &  \frac{\partial L}{\partial t_n}
   = \left[ L , ( L^n )_- \right] , \\
  & & \frac{\partial M}{\partial t_n} = \left[ M , ( L^n )_- \right] +
    n L^{n-1} ,  \\
  & &    \frac{\partial W}{\partial t_n} = - ( L^n)_- W .
\end{eqnarray}
\end{mathletters}

Now we shall show that the $ N $-body Calogero model, in which
the particles  interact through the pairwise inverse square
potentials, share  the same symmetry.
The Hamiltonian of the Calogero model is
\begin{equation}
  {\cal H} = \sum_{j=1}^N p_j^2 + \sum_{1 \leq j < k \leq N}
    \frac{2 a^2}{ (q_j - q_k)^2 } .
\end{equation}
The momentum $ p_j $ and the position $ q_j $ are canonical conjugate,
that is, in terms of the Poisson bracket,
$
 \{ p_j , q_k \} = \delta_{jk} .
$
To discuss the symmetry,  we  use  the classical
$ r $-matrix structure~\cite{AT93} for the $ N \times N $ Lax matrix $
{\rm {\bf L}} $,
\begin{equation}
  {\rm {\bf L}} = \sum_j p_j E_{jj} + \sum_{j \neq k} \frac{{\rm i}a}
  { q_j - q_k }  E_{jk}.
\end{equation}
Here $ E_{jk} $ are the basic matrices, $ ( E_{jk} )_{lm} =
\delta_{jl} \delta_{km} $.
The Lax  matrix $ {\rm {\bf L}} $ satisfies the following identity,
\begin{equation}
  \{ {\rm {\bf L}} \stackrel{\otimes}{,} {\rm {\bf L}} \} =
    \left[ r_{12} , {\rm {\bf L}} \otimes {\rm {\bf 1}} \right] -
    \left[ r_{21} , {\rm {\bf 1}} \otimes {\rm {\bf L}} \right] .
\end{equation}
Here $ \{ \; \stackrel{\otimes}{,} \; \} $ denotes the fundamental
Poisson bracket~\cite{FT87},
$
  \{ {\rm {\bf A}} \stackrel{\otimes}{,} {\rm {\bf B}} \} =
    \sum \{ A_{jk} , B_{lm} \} E_{jk} \otimes E_{lm} .
$
The classical $ r $-matrix is defined by
\begin{equation}
  r_{12} = \sum_{j \neq k} \frac{-1}{q_j - q_k} E_{jk}
    \otimes E_{kj}
    + \sum_{j \neq k} \frac{1}{q_j - q_k} E_{jj} \otimes E_{kj} .
      \label{rmat}
\end{equation}
Using the permutation operator, $ {\rm {\bf P}} = \sum_{j,k} E_{jk}
\otimes E_{kj} $, we may express  $ r_{21} = {\rm {\bf P}} r_{12} {\rm
  {\bf P}} $.
Remark that the classical $ r $-matrix is slightly different
from the one  in~\cite{AT93}.
The other  intriguing structure is possessed by the diagonal matrix $
{\rm {\bf Q}} $,
\begin{equation}
  {\rm {\bf Q}} = {\rm diag} ( q_1, q_2, \cdots, q_N ) .
\end{equation}
This matrix  satisfies the following relation in terms of the classical
$ r $-matrix;
\begin{equation}
  \{ {\rm {\bf Q}} \stackrel{\otimes}{,} {\rm {\bf L}} \}
    = \left[ r_{12} , {\rm {\bf Q}} \otimes {\rm {\bf 1}} \right]
      - {\rm {\bf P}} .
\end{equation}
These equations lead to the following equations of motion;
\begin{mathletters}
\label{lt}
\begin{eqnarray}
  & &  \frac{\partial {\rm {\bf L}}}{\partial t_n}
   = \left[ {\rm {\bf L}} , {\rm {\bf M}}_n \right],    \\
  & &  \frac{\partial {\rm {\bf Q}}}{\partial t_n} = \left[ {\rm {\bf
        Q}} , {\rm {\bf M}}_n \right]
      + n {\rm {\bf L}}^{n-1}  .
\end{eqnarray}
\end{mathletters}
Here $ {\rm {\bf M}}_n $ are $ N \times N $ matrices   defined by
\begin{equation}
  {\rm {\bf M}}_n = n {\rm Tr}_2 \left( r_{12} \cdot
    {\rm {\bf 1}} \otimes {\rm {\bf L}}^{n-1} \right) ,
\end{equation}
where $ {\rm Tr}_2 $ means the trace of the matrices in the second
space of a direct product.
The time variable $ t_n $  is associated with  the flow for the
conserved quantity
$ {\cal I}_n  \equiv {\rm Tr} {\rm {\bf L}}^n $, as
\begin{equation}
  \frac{ \partial {\rm {\bf A}}}{\partial t_n} = \{ {\cal I}_n , {\rm
      {\bf A}} \} .
\end{equation}
One finds that the matrix $ {\rm {\bf M}}_n $ is  equivalent to
the results in~\cite{BR77}.

One observes that (\ref{lt}) have appeared in the KP
hierarchy.
As the time $ t_n $ denote the flow for the conserved quantities
$ {\cal I}_n $,  only the finite quantities are independent for the $
N $-body system.
To compare the Calogero system with the KP hierarchy we must take the
large $ N $ limit, $ N \rightarrow \infty $.
The explicit  construction can be seen from the matrix  representation.
We consider the following equation,
\begin{equation}
  {\rm {\bf Q}} = {\rm {\bf U}} \Bigl( \sum_{k=1}^\infty k t_k {\rm {\bf
          H}}^{k-1}
     \Bigr) {\rm {\bf U}}^{-1} .
   \label{matrix}
\end{equation}
Here we suppose that the matrix $ {\rm {\bf H}} $ does not depend on the
``time'' variables $ t_n $.
A simple calculation leads to (\ref{lt}) if we set
\begin{mathletters}
  \begin{eqnarray}
    & &  {\rm {\bf M}}_n = - \frac{\partial {\rm {\bf U}}}{\partial t_n}
        {\rm {\bf U}}^{-1} ,  \\
    & & {\rm {\bf L}} = {\rm {\bf U H U}}^{-1} .
  \end{eqnarray}
\end{mathletters}
These equations correspond to the Sato equation (\ref{sato}) in the KP
hierarchy.

It is now straightforward to show the relationship between the KP
hierarchy and the  many-body problem from the generalized Lax
equations and the additional symmetry.
The $ n $-th order KP equation corresponds to the flow for $ {\cal I}_n $.
Further  the  KdV hierarchy and the Boussinesq hierarchy, which are the
reduction of the KP hierarchy,  can be  formulated.
For example, the KdV hierarchy is the $ n = 2 $ reduction of
the KP hierarchy.
We  suppose the peudo-differential operator as $ L^2 = \partial^2 + u $,
or $ ( L^2 )_- = 0 $.
Then  the $ k $-th order KdV equation is obtained by
$
  \partial L^2  / \partial t_{2k+1} = \left[ L^2 , ( L^{2k+1} )_-
   \right]  ,
$
while the ``even'' time variables $ t_{2k} $ are suppressed,
$
  \partial L^2 / \partial t_{2k} = 0 .
$
{}From the correspondence with the Calogero model, one concludes that
the $ k $-th order KdV  equation is equivalent to
the flow of $ {\cal I}_{2k+1} $ with the constraint
$
  {\rm grad} {\cal I}_{2k} \equiv 0 .
$
These results  are consistent with the suggestion in \cite{AMM77}.



One of the author (KH) appreciates the  Fellowship through the Japan
Society for Promotion of Science.

\end{document}